\documentclass[12pt]{article}
\usepackage{indentfirst} %% indents the first paragraph
\usepackage{amsfonts,amssymb,amsmath}
\usepackage{bm}          %% For bold math symbols
\usepackage{cite}
\usepackage{url}
\usepackage{enumitem}
\usepackage[usenames,dvipsnames]{xcolor}
\usepackage{todonotes}
\usepackage{multicol}
\usepackage{MnSymbol}
\usepackage{hyperref}
\usepackage{tensor}
\usepackage{tikz-cd}
\usepackage[titletoc,title]{appendix}

\usetikzlibrary{arrows}

\AtBeginDocument{}%

\newcommand{\R}{\mathbb{R}}

\newcommand{\be}{\begin{equation}}
\newcommand{\ee}{\end{equation}}
\newcommand{\bee}{\begin{equation*}}
\newcommand{\eee}{\end{equation*}}
\newcommand{\ba}{\begin{eqnarray}}
\newcommand{\ea}{\end{eqnarray}}
\newcommand{\baa}{\begin{eqnarray*}}
\newcommand{\eaa}{\end{eqnarray*}}
\newcommand{\la}{\label}
\newcommand{\f}{\frac}
\DeclareFontFamily{U}{rsfs}{}         % Formal Script            %
\DeclareFontShape{U}{rsfs}{m}{n}{<5> rsfs5 <6><7> rsfs7          %
  <8><9><10><10.95><12><14.4><17.28><20.74><24.88> rsfs10}{}     %
\DeclareMathAlphabet{\mathfs}{U}{rsfs}{m}{n}                     %
\newcommand{\inter}{{\lrcorner}}
\newcommand{\va}{\scriptscriptstyle}

\newcommand{\cM}{{\cal M}}
\newcommand{\cL}{{\cal L}}
\newcommand{\cH}{{\cal H}}
\newcommand{\cN}{{\cal N}}

\newcommand{\su}{\mathfrak{su}}

\def\i{i}

\def\pb#1{\rlap{\lower1.5ex\hbox{$\longleftarrow$}}{#1}}
\def\dpb#1{\rlap{\lower1.5ex\hbox{$\Longleftarrow$}}{#1}}
\def\spb#1{\rlap{\lower1.5ex\hbox{$\leftarrow$}}{#1}}
\def\sdpb#1{\rlap{\lower1.5ex\hbox{$\Leftarrow$}}{#1}}

\def\ECH{\mathrm{ECH}}

\def\P{\mathrm{P}}
\def\IH{\mathrm{IH}}

\def\D{\Delta}

\def\f{\frac}
\def\rd{\mathrm{d}}

\newcommand{\corurl}{red}
\newcommand{\corcite}{ForestGreen}
\newcommand{\corlink}{blue}

\hypersetup{linktocpage,colorlinks,urlcolor=\corurl,citecolor=\corcite,linkcolor=\corlink,
pdftitle={Black Hole Entropy in Loop Quantum Gravity},
pdfauthor={J.F. Barbero, D. Pranzetti}}

\numberwithin{equation}{section}  %% No need for eqnsection.sty file

\def\QED{{\boldmath$\rule{0.5em}{0.5em}$}}                                % Definimos cómo es el QED (final de la demostración)
\def\markatright#1{\leavevmode\unskip\nobreak\quad\hspace*{\fill}{#1}}    % Definimos: colocar a la derecha ¿no?
\def\qed{\markatright{\QED}}                                              % Definimos el qed final (QED colocado a la derecha

\usepackage{titling}
\usepackage{authblk}
\title{Black Hole Entropy in Loop Quantum Gravity}
\author[1,2]{J. Fernando Barbero G.}
\author[3]{Daniele Pranzetti}
\affil[1]{Instituto de Estructura de la Materia, CSIC. Serrano 123, 28006 Madrid, Spain}
\affil[2]{Grupo de Teor{{\i}}as de Campos y Fisica Estadistica. Instituto Gregorio Mill\'an (UC3M). Unidad Asociada al Instituto de Estructura de la Materia, CSIC}
\affil[2]{Universit\`a degli Studi di Udine, via Palladio 8, 33100 Udine, Italy}

\date{}                     %% if you don't need date to appear
\begin{document}
	
\maketitle
\date{December 27, 2022}

\vspace*{-12ex}

\begin{abstract}
We give an account of the state of the art about black hole entropy in Loop Quantum Gravity. This chapter contains a historical summary and explains how black hole entropy is described by relying on the concept of isolated horizon, with an emphasis on different representations of its associated symmetry group.  It continues with a review of the combinatorial methods necessary to understand the behavior of the entropy as a function of the area and concludes with a discussion of the nature of the quantum horizon degrees of freedom that account for the black hole entropy and the related issue of the fixing of the Immirzi parameter.
\end{abstract}

\section{Introduction}
\label{Intro}

%%%%%%%%%%%   Daniele part  %%%%%%%%%%%%%%

Black holes represent classical solutions of Einstein's equations of General Relativity and they correspond to a final stage of isolated gravitational collapse. Starting on the late sixties, the investigation of black hole (BH) physics has been full of surprises and it revealed intriguing properties which turned these objects into a perfect arena to test any candidate theory of Quantum Gravity. This chapter is  devoted to the application of the Loop Quantum Gravity (LQG) formalism to a statistical mechanical treatment of BH microscopic degrees of freedom (DOF) in order to derive their semi-classical and continuum properties within a quantum gravity approach. The main focus  is the derivation of  the Bekenstein--Hawking entropy formula
\be\la{beke-haw}
S=\frac{k_{\va B}\, A}{4 \ell_{\va P}^2}\,,
\ee
where (in units $c=1$) $k_{\va B}$ is the Boltzmann constant, $\ell_{\va P}=\sqrt{G_{\va N} \hbar}$ is the Planck length and $A$ the BH horizon area.
In fact, right after the analogy between the laws of BH physics and those of ordinary thermodynamics was found in \cite{Bardeen:1973gs}, Bekenstein argued \cite{Bekenstein:1973ur} that a notion of entropy could be associated with a BH, namely $S=\alpha A/\ell_{\va P}^2$, where $\alpha$ is a dimensionless, undetermined constant (from now on we set $k_{\va B}=1$). Plugging this proposal into the BH first law
 \be\la{first}
 \delta M=\frac{\kappa_{\va H}}{8\pi G}\delta A+\Phi_{\va H}\, \delta Q+\Omega_{\va H}\, \delta J,
 \ee
 relating different nearby stationary BH spacetimes, immediately leads to a notion of temperature proportional to the horizon surface gravity $\kappa_{\va H}$, namely $T_{\va H}= \hbar \kappa_{\va H}/(8\pi\alpha)$--- in \eqref{first} $\Phi_{\va H}$ is the
electrostatic potential at the horizon, and $\Omega_{\va H}$ the
angular velocity of the horizon---.

At the classical level though, the temperature of a  BH  is absolute zero since nothing can escape once inside the horizon. By using methods from quantum field theory in curved spacetimes, Hawking arrived at his famous discovery \cite{hawking1975} of BH radiance with a black body spectrum at temperature
\be
 T_{\va H}= \f{\hbar \kappa{\va H}}{2\pi}\,,
 \ee
  thus confirming the physical nature of surface gravity as the temperature of a BH and fixing $\alpha=1/4$, which yields the entropy formula (\ref{beke-haw}).

Although Hawking's derivation relies on a semi-classical regime, analyzing the scattering properties of a quantum test field on the background geometry of a large BH before and after the collapse dynamical phase, the final elegant expression (\ref{beke-haw}) for the BH entropy involves the Planck length, i.e. the scale at which the quantum aspects of the gravitational field cannot be neglected any more. Therefore, a proper statistical mechanical understanding of  the Bekenstein--Hawking formula can only be achieved within a quantum gravity description of the horizon microscopic DOF. It is the recovering of the exact numerical factor in front of the entropy leading term which would then require to go to the semi-classical regime of the given quantum gravity approach. Our goal is to show how these two steps can be completed within the LQG formalism (for other detailed reviews see \cite{DiazPolo:2011np, BarberoG:2015xcq, Perez:2017cmj}).

\subsection{History of black hole entropy in LQG}
\label{sec:BH-LQG}

The beginning of the investigation of BH entropy in LQG can be dated back to the pioneering work of Smolin \cite{Smolin:1995vq}, where an interplay between a topological QFT on the boundary and a non-perturbative quantum gravity description of the gravitational field in the bulk was proposed to describe the state space of a horizon and to provide evidence for  the holographic hypothesis. Subsequently, Rovelli \cite{Rovelli:1996dv} defined a statistical mechanical system modeling a quantum non-rotating and non-charged horizon characterized by a macroscopic parameter given by the LQG eigenvalue of the horizon area. By identifying the BH entropy with the number of microstates of the horizon quantum geometry compatible with the macroscopic configuration, he obtained a result proportional to the area. Combining these results, Krasnov \cite{Krasnov:1996wc} proposed to use $SU(2)$ Chern--Simons (CS) theory to give a  quantum mechanically description for the microscopic states of a  large Schwarzschild BH. The interplay between LQG techniques and the CS formalism of the boundary theory led to a linear dependence of the entropy on the area, with  a proportionality coefficient depending on the Immirzi parameter $\gamma$.

This framework for BH entropy computations in LQG was put on firm ground by the introduction of a quasi-local notion of horizon in equilibrium, with boundary conditions on the gravitational field specified only at a given inner boundary of  spacetime. This is the notion of {\it Isolated Horizon} (IH) introduced in \cite{Hayward:1993wb, Ashtekar:1998sp}  (see also \cite{Ashtekar:2004cn, Booth:2005qc}), which provides a characterization of a static horizon but  eliminates the need to have a knowledge of the complete  spacetime (like in the case of event horizons);  this allows for different bulk dynamical configurations which might be more physically relevant (or even necessary) in certain quantum gravity scenarios. At the same time, the IH framework is restrictive enough to recover the zeroth and first laws of BH thermodynamics \cite{Ashtekar:1999yj, Ashtekar:2000hw}.

In the original analysis of the IH classical phase space \cite{Ashtekar:1999wa}, it was shown that the symplectic structure of general relativity in the first order formalism acquires, through gauge fixing of the internal symmetry, a boundary term parametrized by a $U(1)$ CS connection. Quantization of this enlarged phase space was carried out in \cite{Ashtekar:2000eq}, by respectively applying LQG and CS techniques to the bulk and boundary Hilbert spaces. By coupling the two  through the quantum imposition of IH  boundary conditions for a large, fixed value of the horizon area, the linear behavior in the area of the leading term in the entropy  was confirmed, as well as the need to fix $\gamma$ to a specific numerical value so to recover the Bekenstein--Hawking entropy formula.

While the fixation of the Immirzi parameter through the BH entropy calculation was presented as a way to remove an ambiguity of the quantum formalism, such approach may seem unnatural. In fact, at the classical level  the Immirzi parameter does not encode any physical ambiguity: Different sectors of $\gamma$ simply amount to different canonical transformations in the phase space that do not affect  physical observables. It is only at the quantum level that $\gamma$ becomes a true ambiguity through its appearance in the spectrum of geometrical operators. Therefore, relying on a semi-classical calculation like the one explaining Hawking radiation, where gravity is treated classically, in order to select a specific value of the Immirzi parameter may be an indication that some ingredients are missing---see however Section \ref{sec:DOF} for alternative views on this topic. Nevertheless, the seminal papers \cite{Ashtekar:1999wa, Ashtekar:2000eq} started a rich investigation of the counting problem that led to the discovery of sub-leading logarithmic corrections (independent of $\gamma$) and revealed a discrete structure of the entropy functional
for small values of the IH area \cite{Domagala:2004jt, Meissner:2004ju,  Ghosh:2004wq, Corichi:2006wn, Corichi:2006bs, Agullo:2008yv, FernandoBarbero:2011kb}. It should be noticed that, although the majority of the papers on BH entropy in LQG use the standard area operator, it is actually possible to make use of other natural choices such as the \textit{flux-area operator} \cite{FernandoBarbero:2009ai}. Although qualitatively this works much in the same way as the usual approach, there are some interesting differences. The most obvious one being that the area spectrum consists of equally spaced eigenvalues.

At the same time though, this detailed investigation also led to the emergence of a second conflicting aspect. A logarithmic correction had been found before in \cite{Kaul:2000kf}, without relying on the IH formalism but by counting the conformal blocks of the $SU(2)$ Wess--Zumino--Witten model on a 2-sphere with punctures. A sub-leading term in the entropy was derived shortly  after also by Carlip \cite{Carlip:2000nv}, relying on quite general symmetry considerations about the algebra of constraints in general relativity in the presence of a (local) Killing horizon; the appearance of a natural Virasoro subalgebra motivates the use of  2D CFT methods, previously proposed in \cite{Strominger:1997eq}, which, not surprisingly since in both cases CFT methods play a fundamental role, led to a logarithmic correction with the same numerical coefficient $-3/2$  as found in  \cite{Kaul:2000kf}. Carlip argued that, at least for non-rotating BHs, this coefficient might have a universal nature. However, this expectation was at clash with the result obtained with the $U(1)$ IH framework of \cite{Ashtekar:1999wa, Ashtekar:2000eq}, which yields a numerical coefficient of $-1/2$.

 A way to understand such discrepancy is to observe that the algebra of the $U(1)$ IH quantum boundary conditions, relating the fluctuations of the boundary connection to those of the bulk fluxes defined on surfaces intersecting the horizon, does not preserve the Lie algebraic structure of the classical theory. It follows that, at the quantum level, one can impose only a subset of boundary conditions, leading to a slight overcounting of microstates. This problem with the gauge symmetry reduced model, together with an attempt to provide a more uniform treatment of the bulk and boundary DOF, in order to make contact with the original ideas of \cite{Smolin:1995vq, Rovelli:1996dv, Krasnov:1996wc} and better understand  the role of the Immirzi parameter, led to the development of an $SU(2)$-invariant formulation of IHs \cite{Engle:2009vc, Engle:2010kt,  Perez:2010pq, Frodden:2012en}. These works clarified both the classical and the quantum frameworks, which, in the non-rotating case, allowed to show how the correct imposition of the quantum boundary conditions leads indeed to the $-3/2$ factor of the logarithmic correction \cite{Engle:2011vf}.

 Before reviewing the main technical aspects of the LQG BH entropy calculation, let us point out two  important aspects which are common to  these different technical approaches.
 The first one is the implementation  of  a `weak holographic principle' (see e.g. \cite{Jacobson:1999mi} and  \cite{Markopoulou:1999iq} for an explicit formulation). This dates back to the first works on the subject \cite{Smolin:1995vq, Rovelli:1996dv} (an interesting discussion about it can be found in \cite{Jacobson:2005kr}). Weak holography applied to the BH entropy counting implies that the relevant DOF are only those measurable by observers just outside the horizon. In the LQG literature, this principle has been applied by constructing the horizon density matrix by tracing over all the bulk DOF, both interior and exterior, and assuming the reduced density matrix to be maximally mixed. In this way, only the quantum horizon boundary DOF are considered. These are encoded in the structure of a single intertwiner (either $U(1)$ or $SU(2)$) between all the punctures created by the bulk links piercing the horizon. This intertwiner is assumed to be flat, namely the coarse graining over the bulk DOF is assumed to wash away all the information about the interior bulk curvature (see, however, \cite{Livine:2007sy} for an analysis of the holographic regime of LQG in the presence of bulk entropy) and the total flux across the boundary (in the Ashtekar--Lewandowski vacuum representation \cite{Ashtekar:1993wf}) vanishes. There are valid arguments  \cite{Smolin:2000ag, Rovelli:2017mzl, Perez:2017cmj} why this notion of holography is the only one that can survive in a background independent quantum gravity context, while stronger forms of holography may hold in a fixed background, semi-classical limit.

The second important aspect is that the  LQG intertwiner construction of a quantum IH is purely kinematical. In fact, within the IH framework, it has been shown in \cite{Ashtekar:1999wa} that, in order for the Hamiltonian time evolution (in the covariant phase space formalism) to be well defined, the lapse function smearing  the Hamiltonian constraint of the canonical theory needs to vanish at the horizon. In other words, the bulk dynamics is switched off at the horizon (in this way the horizon area becomes a physical observable) and one assumes that for each IH boundary state there exists at least one physical bulk state compatible with it, which annihilates the Hamiltonian constraint. In this sense the quantum IH construction is purely kinematical.

As a final recent addition to the toolbox used to study BHs in LQG, it is worth mentioning the recent work \cite{Song:2020arr}, where, instead of using a CS theory to describe the quantum DOF sitting at the BH horizon, the authors introduce an $SO(1,1)$ BF-theory. The main consequence of this is the enlargement of the covariant phase space of the system, which can include now spacetime solutions with \textit{any} isolated horizon as inner boundary.

\section{Isolated horizons}\la{sec:IH}

Isolated horizons replace the teleological notion of event horizon with a weaker and local definition involving the behaviour of fields only at the horizon, while allowing us to derive laws analogue to those of BH thermodynamics. Let us review the more relevant features of such a definition as provided in \cite{Ashtekar:1998sp, Ashtekar:1999yj, Ashtekar:2000hw, Ashtekar:2001jb} (for more extensive and technical reviews see, e.~g., \cite{Diaz-Polo:2011uhm, BarberoG:2015xcq}).

%%%%%%%%%%%%%

%Chapter ``Boundary DOF in Loop Quantum Gravity''
%
% Chapter ``Quantum Geometry and Black Holes''
%
%  Chapter ``Loop Quantum Gravity and Quantum Information''

%Chapter ``Emergence of Riemannian Quantum Geometry''

 \subsection{Boundary conditions} \la{sec:BC}

 Let us consider an asymptotically flat  4-manifold $\cM$ with
a metric $g_{ab}$ of signature $(-,+,+,+)$ and  a null hypersurface $\Delta$ of $(\cM, g_{ab})$ with topology $\Delta=S^2 \times \R$. We denote $\ell$ a future-directed {\it null} normal to $\Delta$,  $\nabla_a$ the derivative operator compatible with $g_{ab}$ and  $q_{ab}$ the degenerate intrinsic metric corresponding to the pull-back of $g_{ab}$ to $\Delta$. The IH boundary conditions require, first, that all the field equations and the  stronger dominant energy condition hold at $\Delta$. Furthermore, $\D$ is equipped with an equivalence class $[\ell]$ of null normals, whose members are related by a positive constant rescaling, and  the expansion $\theta_{(\ell)}$ of any given null normal  $\ell\in[\ell]$ has to vanish within $\Delta$. These conditions are enough to guarantee that the horizon area  of a given 2-sphere cross-section $A_s$ is constant in time  (no flux of matter nor gravitational radiation across $\Delta$). Moreover, as implied by the Raychaudhuri equation, the null normal $\ell^a$ is also shear-free, so that the spacetime connection $\nabla_a$ induces a unique intrinsic connection $D_a$  compatible with the induced metric $q_{ab}$. The above conditions imply that the intrinsic metric $q_{ab}$ is Lie dragged by $\ell^a$, that is $\cL_\ell q_{ab}=0$. The final restriction demands that the full intrinsic connection $D_a$ be conserved along $\Delta$, namely $[\cL,D]|_\Delta =0$.   This condition allows one to define a notion of surface gravity $\kappa_\ell$ that is also constant along $\Delta$ for each $\ell\in[\ell]$ without the need to have a Killing field even in the proximity of $\Delta$. We thus recover a generalized zeroth law of BH mechanics.

The isolated horizon boundary conditions provide a generalization of the Killing horizon concept that encompasses all the globally stationary BHs. The freedom left in the positive constant rescaling of the null normal $\ell^a$ reflects on the notion of surface gravity, whose normalization is thus undetermined.  This leads to a family of first laws of  IHs \cite{Ashtekar:2000hw}. In the case of global Killing fields for asymptotically flat spacetimes the same ambiguity is resolved by specifying the normalization of the Killing field at infinity. If required, one can select a unique first law for IHs, e.~g. in the non-rotating case, by matching the surface gravity to that of stationary BHs.

 \subsection{Phase space} \la{sec:PS}

 In order to construct the Hilbert space associated with a quantum IH and thus identify the DOF that account for its entropy, it is crucial to first study the IH phase space structure at the classical level. In order to do so, it proves convenient to employ covariant phase space methods.
 As the LQG quantization framework relies on the holonomy-flux algebra, we consider general relativity it its first order formulation and, more precisely,  we focus on Einstein--Cartan--Holst gravity. In this case, the fundamental variables are represented by a tetrad coframe field composed of   $\mathbb{R}^4$-valued  1-forms $e^I$, with $I=0,i$ and $i=1,2,3$ labeling internal Lorentz indices, and a Lorentz connection $\omega^{IJ}=-\omega^{JI}$ with curvature $F^{IJ}=\rd\omega^{IJ}+\omega^I{}_K\wedge\omega^{KJ}$.

 The Einstein--Cartan--Holst (ECH) Lagrangian is given by
\be\la{ECH Lagrangian}
L_\ECH=\f{1}{2}E_{IJ}\wedge F^{IJ},
\quad
E_{IJ}[e]:=(*+\f1\gamma)(e_I\wedge e_J)\,,
\ee
where the duality operation is defined as $(*M)_{IJ}=\tfrac12{\epsilon_{IJ}}^{KL}M_{KL}$. The time gauge in a ADM-like decomposition of spacetime adopted in canonical LQG is imposed by demanding the component $e^0$ to be a
time-like vector field normal to the Cauchy surface  $\Sigma$ intersecting an IH $\Delta$ at a given 2-sphere cross-section $H=\Delta\cap \Sigma$. Upon this gauge fixing, the symplectic potential of the ECH formulation reads
\be\la{TECH}
 \kappa\,\Theta_\ECH=\frac{1}{\gamma} \int_{\Sigma} E_i\wedge \delta A^i\,,
\ee
 where  $\kappa=16\pi G$,  $E^i:=\epsilon^{i}_{\ jk} e^j\wedge e^k$ and $A^i :=\Gamma^i+\gamma K^i$ is the SU(2) real  Ashtekar connection. Here $K^i:=\omega^{0i}$
 is the extrinsic curvature of $\Sigma$, while $\Gamma^i=-\frac{1}{2}\epsilon^{ijk}\omega_{jk}$ is the spin
connection such that Cartan's equation  $\rd_\Gamma e^i=0$ holds. The  symplectic potential \eqref{TECH} tells us that the gravitational flux $E^i$ and the real  Ashtekar connection $A^i$ are canonical pairs in the bulk $\Sigma$ if we start with the ECH formulation of gravity. However, sticking to a connection variable in the bulk, we see that the  ECH Lagrangian does not give rise to any corner term on the IH cross-section $H$. Such a term would appear if we revert to vector-like variables $(E^i, K^i)$ in the bulk; in fact, using the properties of $\Gamma^i$ it can be shown that \eqref{TECH} can be recast in the form
 \be\la{TECH2}
 \kappa\,\Theta_\ECH
 = \int_{\Sigma} E_i\wedge
\delta K^i+\frac{1}{\gamma} \int_{H} e_i\wedge \delta e^i\,.
\ee
The implications of the corner symplectic potential in \eqref{TECH2} in the quantum theory were studied in \cite{Freidel:2019ees}. If we wish to keep the corner term \textit{and} a connection variable in the bulk, we need to add a boundary Lagrangian term to $L_\ECH$. This yields  the Palatini Lagrangian $L_\P=L_\ECH+\rd \ell$, where the boundary Lagrangian reads \cite{Freidel:2020svx}.%\footnote{The Palatini Lagrangian  can also be written as just a bulk contribution when expressed in terms of the   Ricci scalar $\tilde{R}$ of the slice $\Sigma$ and its extrinsic curvature tensor $\tilde K_{ab}$  \cite{Freidel:2020xyx}.}
 \be
\ell:= \f{1}{2\gamma}  e_I\wedge\rd_\omega e^I\,.
\ee
The reason why we call it Palatini Lagrangian follows from the fact that the associated symplectic potential can be equivalently written as
\begin{equation}\label{TP}
 \kappa\,\Theta_\P=\frac{1}{\gamma}\int_{\Sigma} E_i\wedge\delta A^i+\frac{1}{\gamma} \int_{H} e_i\wedge \delta e^i=\int_{\Sigma} E_i\wedge\delta K^i \,,
\end{equation}
where in the second line we recognize the  familiar canonical symplectic potential expressed in terms of Palatini variables.

 \subsection{Constraints and charges} \la{sec:CC}

 The importance of comparing these two different  formulations of gravity and their associated  symplectic potentials stems from the fact that they lead to symmetry charges for the IH  that can vanish or not according to which formulation one considers. This reflects the difference between {\it gauge} symmetries, which only label gauge redundancies and cannot be used to label  physical states of, say, the IH---since by definition the corresponding  Hamiltonian charges vanish on such states---and {\it physical} symmetries which, instead, possess non-vanishing charges. While this difference does not affect the standard BH entropy calculation in LQG, as reviewed below, it can have important implications for its interpretation regarding the nature of the DOF accounting for it and the role of the Immirzi parameter in the entropy counting (we will come back to this in Sec. \ref{sec:DOF}).

 In the following we concentrate on static isolated horizons, namely we restrict ourselves to the non-rotating case.\footnote{The staticity condition can be formulated by demanding the Newman-Penrose scalar component (in a null tetrad frame adapted to the IH geometry) ${\rm Im}(\Psi_2)$ to vanish \cite{Ashtekar:2001is}.} As already pointed out, in order for the  Hamiltonian time evolution to be well defined, the lapse at the horizon must vanish. Moreover, the time gauge fixes the internal boost freedom. Therefore, we only need to focus on SU(2) internal rotations with parameter $\alpha^i$ and spatial diffeomorphisms generated by  tangent vector fields non-vanishing on the horizon  $v\in T(\Sigma)$. In the bulk these transformations are respectively generated by the Gauss constraint and the vector constraint, respectively
 \be
 \rd_A E^i=0\,,
\quad
(v\inter F^i(A))\wedge E^i=0\,.
\ee
By plugging the corresponding field transformations $\delta_\alpha, \delta_v$ into the symplectic form $\Omega = \delta \Theta$, on-shell of these constraints (denoted with  $\hat{=}$), it is easy to show that \cite{Engle:2010kt}
\be
\kappa\Omega_\P(\delta_\alpha, \delta)~\hat{=}~0\,,
\quad
\kappa\Omega_\P(\delta_v, \delta)~\hat{=} \int_H \delta\left((v\inter E_i) \wedge K^i\right)= 0\,,
\ee
where the last equality holds due to the boundary conditions for static isolated horizons.  Therefore, in the Palatini formulation  defined by the symplectic potential \eqref{TP}, SU(2) rotations and tangent diffeomorphisms represent degenerate directions of the symplectic form for a static IH and their corresponding Hamiltonian  charges  vanish. Stated otherwise,  in this case the   IH symmetry group
\be\la{GH}
G_H= {\rm Diff}(H)\ltimes SU(2)^H
\ee
is trivially represented at the classical level for the static case.

On the other hand, if we consider the ECH formulation defined by the symplectic potential \eqref{TECH}, one can show  \cite{Freidel:2020svx}
\be\la{QECG}
\kappa\Omega_\ECH(\delta_\alpha, \delta)~\hat{=}~\f1\gamma \int_H  \delta( \alpha^i E_i)\,,
\quad
\kappa\Omega_\ECH(\delta_v, \delta)~\hat{=}
% ~\f1\gamma\int_H \delta(v\inter E_i \wedge \Gamma^i)
-\f2\gamma\int_H \delta(v\inter e_i \rd e^i)
\,.
\ee
Hence, in this case the IH symmetry group \eqref{GH} has a canonical representation in the gravitational phase space of an isolated horizon.
In particular, it is straightforward to verify that the first set of charges in \eqref{QECG}, $E[\alpha]=1/{(\kappa\gamma)} \int_H   \alpha^i E_i$,  reproduces the non-commutativity relation of the LQG fluxes already at the classical and continuum level, as  pointed out already in \cite{Engle:2010kt}. More precisely, by means of the corner symplectic potential expressing the non-commutativity of the frame field on $H$ we recover the $\su(2)$ Lie algebra
\be
\{E[\alpha],E[\beta]\}= E\big[[\alpha,\beta]\big]\,.
\ee
These non-vanishing charges generate physical symmetries and  can be used to label different geometrical states of the horizon.
The  infinite dimensional IH symmetry group \eqref{GH} is a subgroup of the {\it corner symmetry group} representing the universal  maximally extended  subgroup of bulk diffeomorphism in the presence of an embedded  codimension-2 surface. We refer the reader to the Chapter ``Boundary Degrees of Freedom in Loop Quantum Gravity'' for a review of this subject and references therein.
Also, a  recent study of the symmetry group and charges of non-expanding horizons has been carried out in \cite{Ashtekar:2021wld,Ashtekar:2021kqj}.

Finally, the passage to a connection parametrization of the IH corner phase space, which led to the original CS description of the boundary theory, can be achieved by means of a horizon constraint relating the curvature of the real Ashtekar connection at $H$  to the corner flux. For  spherical IHs  considered in the rest of the Chapter, this constraint takes the form
\be\la{FE}
F^i(A) \stackrel{H}= -\frac{\pi (1-\gamma^2)}{A_{H}}\, {E}^{i}\,.
\ee
Using this constraint, the symplectic form of spherical IHs in the Palatini formulation can be recast as \cite{Engle:2010kt}
\be\la{SF}
\kappa\,\Omega_\P(\delta,\delta) =\f1\gamma \int_\Sigma
\delta E^i\wedge \delta A_i- {\frac{A_{ H}}{\pi \gamma
({1-\gamma^2})}} \int_H \!\!\! \delta A ^i\wedge
\delta A_i \,,
\ee
where the corner term now reproduces the symplectic structure of an SU(2) Chern--Simons  theory at level
\be\la{k}
k_{CS}=
\frac{A_H}{4\pi \ell_p^2\gamma (1-\gamma^2)}\,.
\ee
The symplectic form \eqref{SF} together with  \eqref{FE} represent the starting point for the original IH quantization yielding the single intertwiner model introduced below. %We will review the basic ingredients of this quantum construction in the next section.

As reviewed in  Section \ref{sec:BH-LQG}, the CS parametrization of the IH corner phase space was initially revealed  in a U(1) gauge fixed formulation \cite{Ashtekar:1997yu,Ashtekar:2000eq}. The inclusion of distortion for static isolated horizons was introduced in both the SU(2) and U(1) constructions respectively in \cite{Perez:2010pq} and \cite{Beetle:2010rd}; while progress towards the inclusion of rotation was made in \cite{Ashtekar:2004nd, Frodden:2012en, BenAchour:2016mnn, Gambini:2018ucf, Gambini:2020fnd}.
Alternative parametrizations involving BF-like variables and amenable to 2+1 LQG quantization techniques were proposed in \cite{Wang:2014oua,Pranzetti:2014tla, Huang:2015cma}.
A generalization to higher dimensional horizons and supersymmetry  was formulated in \cite{Lewandowski:2004sh,Bodendorfer:2013jba, Eder:2022gge}.
The addition of gauge charges was studied originally in \cite{Ashtekar:1999wa} and later more thoroughly in \cite{Eder:2018uzm}.
The extension to  topologies different than the spherical one was considered in \cite{DeBenedictis:2011hh}.

\subsection{Quantum geometry}

The symplectic structure \eqref{SF} consists of a bulk and a corner contribution. The bulk term, parametrized by the real  Ashtekar connection and its conjugate flux, lends itself to a quantization in terms of  standard LQG techniques. This yields a bulk Hilbert space with an orthonormal basis of quantum geometry states spanned by spin networks. These are states labelled by: A collection of links joining at nodes of arbitrary valence and forming a closed graph, a semi-integer  positive number $j$ (spin)--unitary irreducible representation of SU(2)--assigned to each link, and an invariant tensor (intertwiner)  in the tensor product of SU(2) representations labelling the links converging at the given node.
We refer the reader to the Chapter ``Emergence of Riemannian Quantum Geometry'' for more details on the construction of the LQG Hilbert space of quantum geometry.

In the presence of the spacetime inner boundary associated with the IH, some of the links can pierce the horizon 2-sphere. From an outside observer point of view, such links end at the horizon, where they create {\it punctures} labelled by both the link spin $j$ and its corresponding magnetic number $m$. This way, the bulk Hilbert space can be represented as the orthogonal sum of open spin networks with one link ending at each of the $n$ points on the horizon forming a given finite set $P$ with $\{j\}_n=\{j_1,\dots, j_n\}$, namely
\be
\cH_\Sigma=\bigoplus_{P,\{j\}_n} \cH^{P,\{j\}_n}_\Sigma\,,
\ee
such that all states in each subspace $\cH^{P,\{j\}_n}_\Sigma$ yield the same horizon area eigenvalue. This decomposition is useful for the entropy counting in the area ensemble (Sec. \ref{sec:comb}).

The fact that the punctures are labelled also by (in general) non-vanishing magnetic numbers indicates that, in the quantum theory, local SU(2) charges defined on a small patch of the horizon around each puncture are different from zero. From the perspective of the Chern--Simons theory, these charges can be understood, by means of the IH constraints \eqref{FE}, as sources of conical curvature singularities. In order to restore {\it local} SU(2) gauge invariance at the punctures, one then needs to add corner DOF, which can be understood as topological defects sourcing a distributional curvature for the Chern--Simons theory. It is important to stress that, in this formulation, these so-called would-be-gauge DOF \cite{Carlip:1998wz} have a purely quantum origin. This way, the corner Hilbert space can be identified  with that of a Chern--Simons theory  on a punctured 2-sphere with level $k_{CS}$ given by \eqref{k}, with flat curvature everywhere except at the location of the punctures in a given ordered set $P$ \cite{Ashtekar:2000eq,Engle:2009vc}. Upon identifying the Chern--Simons punctures DOF with the bulk spin labels, as a consequence of the quantum imposition of \eqref{FE}---which restores local gauge invariance---the total Hilbert space in the presence of a quantum IH can be written as
\be
\cH_{\IH}=\bigoplus_{P,\{j\}_n} \cH^{P,\{j\}_n}_\Sigma\otimes \cH^{P,\{j\}_n}_{H}\,.
\ee
Up to this point, the construction of the IH Hilbert space has been purely kinematical. In order to define the physical Hilbert space, spatial diffeomorphism invariance needs to be implemented as well.\footnote{Recall that the Hamiltonian constraint does not play a direct role, as the lapse smearing function needs to vanish on the horizon.} This is a subtle but important aspect of the entropy counting. As pointed out at the beginning of Section \ref{sec:CC}, if the tangent diffeomorphism charges vanish at the horizon, as it does in the Palatini formulation \eqref{SF}, then the position of punctures on the horizon cannot be regarded as a physical quantity. This means that states corresponding to  different  localizations of the punctures on the horizon need to be considered as physically equivalent, and only the number $n$ of punctures is required to characterize physical states. At the same time, in order for the quantization of the corner phase space to be well defined, an ordering of the punctures needs to be introduced. Moreover, for the entropy counting to yield a result linear in the horizon area, the punctures need to be considered {\it distinguishable},\footnote{See \cite{Ghosh:2013iwa, Pithis:2014uva} for the analysis and alternative models involving different statistics of the punctures.} so that different orderings count as different physical states and contribute to the entropy---the importance of distinguishability between two sets of punctures differing by their ordering was originally pointed out by Krasnov \cite{Krasnov:1996tb}---.  Different orderings can be obtained by the action of tangent diffeomorphisms, which act transitively on this additional structure. This means that, in the quantum theory, a subset of diffeomorphism charges at the horizon needs to be activated, the same way as a finite set of local SU(2) charges are. In the standard treatment then, also these extra and necessary diffeo DOF are considered to have a purely quantum origin---we will come back to this important interpretational aspect in Sec. \ref{sec:DOF}.

Hence, upon imposition of the spatial diffeomorphism constraint, and keeping in mind the distinguishable statistical character of the punctures, the IH physical Hilbert space for given number $n$ of punctures can be written as
\be
\cH^{n}_{\IH}=\bigoplus_{\{j\}_n} \cH^{\{j\}_n}_\Sigma\otimes \cH_H^{CS}(j_1,\dots,j_n)\,,
\ee
where the spins $j_1,\dots,j_n$ are subject to the horizon area constraint
\be\la{AC}
A_H-\delta\leq 4\pi\gamma \ell_P^2\sum_{i=1}^n\sqrt{j_i(j_i+1)} \leq A_H+\delta\,,
\ee
where we introduced an area interval  with $\delta$ of the order of the Planck area.
Finally, there is an additional global constraint that follows from \eqref{FE} and the spherical topology of the horizon, which implies that a loop going around all the punctures is contractible and, hence, the holonomy along it should be trivial. This implies the inclusion
\be\la{HSpher}
 \cH_H^{CS}(j_1\cdots  j_n)
\subset{\rm Inv}(j_1\otimes\cdots \otimes j_n)\,,
\ee
where ${\rm Inv}$ denotes the invariant subpace in the tensor product, which becomes  an equality in the large area limit $A_H\propto k_{CS}\to\infty$. Therefore, for a large BH, the horizon Hilbert space can be effectively identified with a SU(2) intertwiner space between the spins associated to all the punctures. This single intertwiner picture can also be understood as the result of a coarse graining procedure of the DOF of a spin network in the BH interior \cite{Livine:2006xk}.

By further tracing over the exterior bulk DOF while restricting to horizon states  compatible with \eqref{AC}, in accordance with the weak holographic principle advocated above, we arrive at the IH density matrix $\rho_\IH$. Demanding the final BH state to be a a maximally mixed state, or equivalently the validity of a maximal entropy principle---which is  expected to capture some relevant features of the effective dynamics in the continuum limit \cite{Oriti:2015rwa}---, the  quantum statistical mechanical horizon entropy is given by
\be\la{S}
S=-Tr(\rho_{\IH}\ln\rho_{\IH})=\ln(\cN_\IH)\,,
\ee
where $\cN_\IH$ is the dimension of the IH corner Hilbert space. Its derivation through  combinatorial methods is the subject of  the next section.

\section{Black holes and combinatorics}\la{sec:comb}

In order to compute the entropy of a physical system one has to count the number of microscopic configurations compatible with its macroscopic state, i.e. solve a \textit{combinatorics} problem. Ideally, one would like to have closed expressions for the entropy, but often, it is necessary to work with asymptotic expansions to really understand the behavior of the entropy in the large area limit (for which we expect that the ``static'' description discussed above is good enough).

The specific nature of the combinatorial problems relevant to the computation of BH entropy makes it necessary to work with diophantine equations. Two other technical tools are also useful: generating functions and Laplace transforms (as first pointed out in \cite{Meissner:2004ju}). As we will show, from the Laplace transform of the BH entropy as a function of its area, it is possible to obtain the large area behavior. This is how the Hawking law is recovered in this setting. In the context of LQG, the use of combinatorial methods shows up at a very basic level; for instance, when trying to understand the area operator. The distribution of area eigenvalues can be studied in great detail \cite{FernandoBarberoG:2017mlp} by using methods similar to those employed to computate of the entropy, so we will give a short account on this problem in the following.

\subsection{The spectrum of the area operator}\label{spectrum area}

The spectrum of the area operator $\hat{A}_S$ associated with a surface $S$ has a complicated structure (see \cite{Ashtekar:1996eg,Ashtekar:2004eh}), however, when studying BHs in LQG modelled with the help of isolated horizons, the relevant part of it is given by
\begin{equation}\label{BH_area_spectrum}
A_s=4\pi\gamma \ell_{\va P}^2\sum_{j=1}^n\sqrt{k_j(k_j+2)}\,;\quad k_j\in \mathbb{N}\,;\quad n\in\mathbb{N}\,.
\end{equation}
For simplicity, in the following we will use units such that $4\pi\gamma \ell_{\va P}^2=1$. The combinatorial problem that must be solved in order to describe the distribution of the area eigenvalues can be phrased as follows:

\textit{For every positive number $a>0$ determine the function $N(a)$ defined as one plus the number of different multisets consisting of positive integers $k_j\in \mathbb{N}$ such that }
\[
\sum_{j\in \mathbb{N}}\sqrt{k_j(k_j+2)}\leq a\,.
\]

The value of $N(a)$ tells us the number of eigenvalues of the area in the interval $[0,a]$ taking into account the degeneracy associated with the fact that different multisets of positive integers may give the same area eigenvalue. Several approximate ways to study this problem have been discussed in the literature (see \cite{FernandoBarberoG:2017mlp} and references therein), however, it is possible to tackle it without relying on popular, but difficult-to-control, approximations such as $\sqrt{k_j(k_j+2)}\sim k_j$ and $\sqrt{k_j(k_j+2)}\sim k_j+1$. Notice, by the way, that $N(a)$ is a \emph{staircase function}, i.e. an increasing function which is constant except at a countable set of values of the independent variable $a$ where it jumps (\textit{precisely} the eigenvalues of the area operator!). The Laplace transform\footnote{If $f:\mathbb{R}^+\rightarrow \mathbb{R}:x\mapsto f(x)$, we  denote its Laplace transform as $\mathcal{L}(f;s):=\int_0^\infty e^{-sx}f(x)\mathrm{d}x$.} of functions of this type can often be obtained in closed form because the Heaviside $\theta(a-a_0)$  step function ($a_0>0$) can be expressed as\footnote{The integration contour, denoted with the limits $c-i\infty$ and $c+i\infty$, is the straight line, parallel to the imaginary axis, $\mathrm{Re}(z)=c$ with $c$ larger than the real part of the singularity in the integrand.}
\[
\theta(a-a_0)=\frac{1}{2\pi i}\int_{c-i \infty}^{c+i\infty}\frac{e^{(a-a_0)s}}{s}\mathrm{d}s\,, \quad (c>0)\,,
\]
and, hence, if the jumps of $N(a)$ at $a_n>0$ ($n\in \mathbb{N}$) are $\beta_n$, we can write
\begin{equation}\label{N(a)}
N(a)=\frac{1}{2\pi i}\int_{c-i \infty}^{c+i\infty}\frac{e^{as}}{s}\widehat{N}(s)\,,\quad\mathrm{with}\quad \widehat{N}(s):=\sum_n\beta_n e^{-a_n s}\,.
\end{equation}
In practice, the effectiveness of this strategy hinges on the possibility to write the function $\widehat{N}(s)$ in an appropriate closed form. As discussed in \cite{BarberoG:2008dwr,BarberoG:2008dee}, by using generating functions and some properties of the solutions to the Pell equation, it is actually possible to find the following expression for $\widehat{N}(s)$
\begin{equation}\label{hatN(s)}
\widehat{N}(s)=\prod_{k=1}^\infty \frac{1}{1-\exp\big(-s\sqrt{k(k+2)}\,\big)}\,.
\end{equation}
Plugging this into \eqref{N(a)} provides an integral representation for $N(a)$ which can be used to extract the asymptotic behavior of $N(a)$ when $a\rightarrow\infty$, and approximate expressions in other regimes (for instance, for small values of $a$) \cite{FernandoBarberoG:2017mlp}.

As the derivation of \eqref{hatN(s)} is actually the first step in the BH entropy computations, we will give now some details about it. An important preliminary comment is that the eigenvalues of the area \eqref{BH_area_spectrum} can always be written as linear combinations, with integer coefficients, of square roots of square free numbers\footnote{These are positive intergers which can be written as products of different primes, i.e. such that their prime number decomposition has no repeated factors.} $p_j$ ($j\in \mathbb{N}$),  $(p_1=2\,,p_2=3\,,p_3=5\,,p_4=6\,,\ldots)$. If a certain linear combination $\tilde{a}:=\sum_{k=1}^r q_k\sqrt{p_k}$ is to be an eigenvalue of the area operator, it must be possible to find solutions to the equation
\begin{equation}\label{Eq1}
\sum_{k\in \mathbb{N}}n_k\sqrt{k(k+2)}=\sum_{j=1}^r q_j\sqrt{p_j}
\end{equation}
in the unknowns $k\in\mathbb{N}$ and $n_k\in\mathbb{N}$. A solution $(k,n_k)$ means that a state with $n_k$ edges piercing the horizon and carrying $k/2$ spin labels is an eigenfunction of the area operator with eigenvalue $\tilde{a}$. If several solutions exist, the eigenvalue is degenerate and, if no solution exists, then $\tilde{a}$ does not belong to the spectrum of the area operator.
Each term of the form $\sqrt{k(k+2)}$ in \eqref{Eq1} can be written as the product of an integer times the square root of a squarefree number (SRSFN), hence, the left hand side of \eqref{Eq1} will always be a linear combination of SRSFN's with coefficients given by integer linear combinations of the $n_k$. Then, we have to identify, for each of the $p_j$ appearing in $\tilde{a}$, the possible values of $k$ satisfying
$
\sqrt{k(k+2)}=\sqrt{(k+1)^2-1}=y\sqrt{p_j}
$
for some $y\in \mathbb{N}$. This amounts to solving the \textit{Pell equation}
\begin{equation}\label{Pell}
(k+1)^2-p_j y^2=1
\end{equation}
in the unknowns $k$ and $y$ for each $p_j$. The solutions to \eqref{Pell} are well known. For each squarefree $p_j$ they are an infinite set of the form $(k^j_m,y^j_m)$ labeled by $m\in \mathbb{N}$. They all derive from a \textit{fundamental solution} $(k^i_1,y^i_1)$---which corresponds to the lowest values of $k$ and $y$---and are given by a simple formula (see  \cite{Agullo:2008yv}). For instance, for $p_1=2$ the fundamental solution to the Pell equation \eqref{Pell} is $(2,2)$ and the first solutions are $(2,2)\,,(16,12)\,,(98,70)\,,\ldots$

By doing this for all the square free numbers $p_i$ appearing in $\tilde{a}$ we can put the left hand side of \eqref{Eq1} in the form
$
\sum_{j=1}^r\sum_{m=1}^\infty n_{k^j_m}y^j_m\sqrt{p_j}
$,
with $n_{k^j_m}$ non-negative integers, and write \eqref{Eq1} as
\begin{equation}\label{Eq2}
 \sum_{j=1}^r\sum_{m=1}^\infty n_{k^j_m}y^j_m\sqrt{p_j}=\sum_{j=1}^r q_j\sqrt{p_j}\,.
\end{equation}
Taking into account that the $\sqrt{p_j}$ are linearly independent over the rationals, equation \eqref{Eq2} can actually be written as the system of $r$ equations
\begin{equation}\label{Eq3}
\sum_{m=1}^\infty y_m^j n_{k^j_m}=q_j\,,\quad j=1,\ldots,r
\end{equation}
in the unknowns $n_{k^j_m}$. It should be noted that the sum in \eqref{Eq3} is always finite with a number of terms that depends on $q_j$ (the $y_m^j$ grow with $m$). Another important fact is that, for different square free numbers $p_{j_1}$ and $p_{j_2}$, the sets $\{k^{j_1}_m:m\in \mathbb{N}\}$ and $\{k^{j_2}_m:m\in \mathbb{N}\}$ \textit{are always disjoint}. This can be shown by noting that if $k\in \{k^{j_1}_m:m\in \mathbb{N}\}\cap \{k^{j_2}_m:m\in \mathbb{N}\}$, then, there exist positive integers $y_1$ and $y_2$ such that $p_1/p_2=y_1^2/y_2^2$, but this is impossible. Indeed, the irreducible form of the fraction $p_1/p_2$ is the quotient of two square free numbers (say, $\pi_1$ and $\pi_2$) and the irreducible form of the fraction $y_1^2/y_2^2$ is the quotient of the squares of two integers (say, $z_1$ and $z_2$). Now, the irreducible form of a fraction is unique, hence, $\pi_1=z_1^2$ and $\pi_2=z_2^2$, which leads to a contradiction as $\pi_1$ and $\pi_2$ are square free.

We then see that the variables $n_{k^j_m}$ appearing in each of the equations in \eqref{Eq3} are \textit{different}. As a consequence, the equations are independent and can be solved separately. There are, in fact, well-known algorithms to do this. However, for our purposes  it suffices to know the number of solutions to these equations. This can be easily achieved by using generating functions. For instance, let us consider the following equation in the non-negative, integer unknowns $z_j$, $j\in\mathbb{N}$ (here, $a_j\in\mathbb{N}$)
\[
\sum_{j=1}^N a_jz_j=q
\]
for each $q\in \mathbb{N}$. Now, the number of solutions to this equation is the coefficient of the $x^q$ term in the Taylor expansion about $x=0$ of the function
\[
f(x)=\frac{1}{\prod_{j=1}^N(1-x^{a_j})}
\]
---that we denote as $[x^q]f(x)$---, as can be easily seen by multiplying the Taylor expansions of $1/(1-x^{a_j})$, and tracking the origin of the terms that add up to give each power of $x$ in the expansion. By proceeding in this way, it is straightforward to see that the number of solutions to \eqref{Eq2} can be written as $[x_1^{q_1}\cdots x_r^{q_r}]G(x_1,x_2,\ldots)$ with $G$ given by the following formal series involving an infinite number of variables $x_j$, each one of them associated with the corresponding square free integer $p_j$,
\begin{equation}\label{GeneratingF1}
G(x_1,x_2,\ldots)=\prod_{j=1}^\infty\prod_{m=1}^\infty \frac{1}{1-x_j^{y^j_m}}\,.
\end{equation}
Notice that, for a concrete choice of a finite number of $q_j$, one only needs to consider a truncation of $G$ involving a finite number of variables---the ones associated with the square free numbers appearing in the right hand side of \eqref{Eq1}---. Also, given a particular value of $\tilde{a}$, the maximum value of each $y^j_m$ is bounded by $\tilde{a}$; this constraints the possible values of $m$. Now, we are in the position to finally obtain \eqref{hatN(s)}. In order to do so, we simply have to turn the terms of the form $x_1^{q_1}\cdots x_r^{q_r}$ in the power series expansion of $G$ into $\exp\big(-s\sum_{j=1}^r q_j\sqrt{p_j}\big)$. This can easily be achieved by replacing each $x_j$ by $\exp(-s\sqrt{p_j})$. This leads to
\[
\widehat{N}(s)=G\big(e^{-s\sqrt{p_1}},e^{-s\sqrt{p_2}},\ldots\big)=\prod_{j=1}^\infty \prod_{m=1}^\infty\frac{1}{1-e^{-sy^j_m\sqrt{p_j}}}=\prod_{j=1}^\infty \prod_{m=1}^\infty\frac{1}{1-e^{-s\sqrt{k^j_m(k^j_m+2)}}}\,.
\]

\vspace*{-3mm}

To end, we notice that, as $\{k^{j_1}_m:m\in \mathbb{N}\}$ and $\{k^{j_2}_m:m\in \mathbb{N}\}$ are always disjoint, and every $k\in \mathbb{N}$ is the solution to some Pell equation (corresponding to a square free number that can be identified by computing $\sqrt{k(k+2)}$ and taking out from the square root as many factors as possible), the products appearing in the preceding expression can be written as in \eqref{hatN(s)}.

Several comments are in order now:

\begin{itemize}
  \item The expression for $N(a)$ given by (\ref{N(a)},\ref{hatN(s)}) is very well suited to analyze the spectrum of the area operator \eqref{BH_area_spectrum} because it encodes both the position of the area eigenvalues and the degeneracy associated with the fact that the same area eigenvalue can correspond to different spin network states.
  \item It should be pointed out, however, that the number of area eigenstates smaller or equal than a given area $a$ is given by $\lim_{A\rightarrow a^+ N(a)}$ and not by $N(a)$. This is so because the value of the inverse Laplace transform at a jump singularity is the average of the left and right limits there.
  \item It is important to realize that, if the area spectrum were equally spaced, it would be possible to encode the content of the function $N(a)$ in a (formal) power series of a single variable. In this sense, Laplace transforms prove to be far superior because they can accomodate more general situations as the one relevant here.
\end{itemize}

\subsection{Black hole entropy computations}\label{BHentropy}

In the preceding subsection we have looked at the spectrum of the area operator. We will briefly explain now how the BH entropy is defined according to some prescriptions considered in the literature\footnote{We will not be exhaustive here, we will just consider a particular example which, nonetheless, we consider as sufficiently illustrative.} and how it can be computed by using generating functions along the lines spelled above. Here we will focus on the $U(1)$ case; we will make some comments on the $SU(2)$ case at the end of the section. A typical phrasing of the counting problem that must be solved in order to compute the entropy of a BH is the following \cite{Domagala:2004jt}

\textit{The entropy $S(a)$ of a BH of area $a$ is $\log\big(1+N(a)\big)$, where $N(a)$ is the number of all the arbitrarily long, finite, sequences $(k_1,\ldots,k_n)$ of non-zero integers such that the following two conditions hold:
\[
\sum_{j=1}^N\sqrt{|k_j|(|k_j|+2)}\leq a\,,\quad \sum_{j=1}^N k_j=0\,.
\]}

The problem can be solved by following these four steps:
\begin{enumerate}
  \item For a given value $\tilde{a}$ of the area find the number of ways to choose positive integers $|k_j|\in \mathbb{N}$ such that
  \begin{equation}\label{condition1}
  \sum_{j=1}^N\sqrt{|k_j|(|k_j|+2)}=\tilde{a}\,.
  \end{equation}
        At this stage we do not care about order, i.e. we only need to find out how many times each integer appears (we just count \textit{multisets}).
  \item Count the possible ways of ordering the multisets obtained in step 1.
  \item Count all the ways to introduce signs in the sequences of integers considered in step 2 in such a way that the condition $\sum_{j=1}^Nk_j=0$ holds. Here $k_j$ refers to $|k_j|$ with a positive or negative sign.
  \item Repeat for all the area eigenvalues smaller or equal than $\tilde{a}$ and add the results.
\end{enumerate}

\noindent \textbf{Step 1} has been essentially solved in subsection \ref{spectrum area} where we discussed how to count the number of solutions to the diophantine equations that tell us the different ways to get a given area eigenvalue. As we showed, a neat way to encode this information was to use the generating function \eqref{GeneratingF1}. In this step we just determine the number of multisets (configurations) $\{(k_m^j,n_{k_m^j})\}$ associated with a given $\tilde{a}=\sum_{j}q_j\sqrt{p_j}$.

\noindent \textbf{Step 2}, let us consider a configuration
\[
\Big(\underbrace{1,\ldots,1}_{n_1},\ldots, \underbrace{k,\ldots,k}_{n_k},\ldots, \underbrace{k_{\rm max},\ldots,k_{\rm max}}_{n_{k_{\rm max}}}\Big)\,,\quad n_1\,,n_2\,,\ldots\,,n_{k_{\rm max}}>0\,.
\]
The number of ways to reorder the elements in this configuration is just given by the multinomial coefficient
$
{\left(\sum_{k=1}^{k_{\mathrm{max}}}n_k\right)!}/{\prod_{k=1}^{k_{\mathrm{max}}}n_k!}
$.

We can now modify the generating function \eqref{GeneratingF1} in such a way that the coefficient of each term gives us the number of possible reorderings. The way to do this is explained in \cite{BarberoG:2008dee}. The final result is
\begin{equation}\label{GeneratingF2}
G^{(2)}(x_1,x_2,\ldots)=\left(1-\sum_{j=1}^\infty\sum_{m=1}^\infty x_j^{y^j_m}\right)^{-1}\,.
\end{equation}

\noindent \textbf{Step 3} The condition $\sum_{j=1}^Nk_j=0$ (often referred to as the \textit{projection constraint}) can be taken into account by including an extra variable in  \eqref{GeneratingF2}. As discussed in \cite{BarberoG:2008dee} and \cite{Agullo:2010zz} the generating function
\begin{equation}\label{GeneratingF3}
G^{\mathrm{DL}}(z,x_1,x_2,\ldots)=\frac{1}{1-\sum_{j=1}^\infty\sum_{m=1}^\infty (z^{k^j_m}+z^{-k^j_m})x_j^{y^j_m}}
\end{equation}
(here the label $\mathrm{DL}$ stands for Domagala--Lewandowski) is such that
\[
[x_1^{q_1}x_2^{q_2}\cdots][z^0]G^{\mathrm{DL}}(z,x_1,x_2,\ldots)=:[x_1^{q_1}x_2^{q_2}\cdots]\widetilde{G}^{\mathrm{DL}}(x_1,x_2,\ldots)
\]
gives the number of sequences of integers $(k_1,\ldots,k_N)$ satisfying the conditions
\[
\sum_{j=1}^N\sqrt{|k_j|(|k_j|+2)}=a\,,\quad \sum_{j=1}^N k_j=0\,.
\]
We will write this number as $D^{\mathrm{DL}}(q_1\sqrt{p_1}+q_2\sqrt{p_2}+\cdots)$ and refer to it as the \textit{BH degeneracy} associated with the area eigenvalue $q_1\sqrt{p_1}+q_2\sqrt{p_2}+\cdots$. Notice that $D^{\mathrm{DL}}(0)=1$. This will take care of the $1$ introduced in the definition of the entropy as $\log\big(1+N(a)\big)$.
By using Cauchy's theorem we can write
\[
[z^0]G^{\mathrm{DL}}(z,x_1,\ldots)=\frac{1}{2\pi i}\oint_c\frac{\mathrm{d}z}{z} G^{\mathrm{DL}}(z,x_1,\ldots)=\frac{1}{2\pi}\int_0^{2\pi}\mathrm{d}\theta G^{\mathrm{DL}}(e^{i\theta},x_1,\ldots)\,,
\]
where $c$ is a positively oriented contour around the origin that we parameterize as $z=e^{i\theta}$. For \eqref{GeneratingF3} this gives
\begin{equation}\label{GeneratingF4}
\widetilde{G}^{\mathrm{DL}}(x_1,x_2,\ldots)=\frac{1}{2\pi}\int_0^{2\pi}\frac{1}{1-2\sum_{j=1}^\infty\sum_{m=1}^\infty \cos(k_m^j\theta)x_j^{y^j_m}}\mathrm{d}\theta\,.
\end{equation}

\noindent \textbf{Step 4} To conclude, in order to compute the entropy $S(a)$ we have to take into account the inequality that appears in its definition. To this end, given a value $a$ of the area, we have to repeat the previous procedure for all the area eigenvalues, smaller or equal to $a$, and add the resulting BH degeneracies.

If we had a concrete formula giving the eigenvalues $a_n$ of the area operator as a function of $n\in\mathbb{N}$, this task would indeed be very simple: we would first build the generating function $f(z)=\sum_{n=0}^\infty D^{\mathrm{DL}}(a_n)z^n$ (maybe as a formal power series) and then consider $f(z)/(1-z)$ (see \cite{Sahlmann:2007jt}). However, as far as we know, no such formula is available. The alternative is to use Laplace transforms as shown in the following. To begin with we can write
\[
\sum_{\{n:a_n\leq a\}}D^{\mathrm{DL}}(a_n)=\int_0^a \sum_{n=1}^\infty D^{\mathrm{DL}}(a_n)\delta(a'-a_n)\mathrm{d}a'\,.
\]
Remembering that
$
\mathcal{L}(\delta_{a_0},s)=e^{-a_0s}\,, (a_0>0)\,, \mathcal{L}\left(\int_0^af(a')\mathrm{d}a',s\right)=\frac{1}{s}\mathcal{L}(f,s)
$,
%\begin{align*}
%  &\mathcal{L}(\delta_{a_0},s)=e^{-a_0s}\,,\quad (a_0>0)\,, \\
%  &\mathcal{L}\left(\int_0^af(a')\mathrm{d}a',s\right)=\frac{1}{s}\mathcal{L}(f,s)\,,
%\end{align*}
we have
\[
\sum_{\{n:a_n\leq a\}}D^{\mathrm{DL}}(a_n)=\mathcal{L}^{-1}\left(\frac{1}{s}\sum_{n=1}^\infty D^{\mathrm{DL}}(a_n)e^{-a_ns},a\right)\,.
\]
The key insight now is
\begin{align*}
\sum_{n=1}^\infty D^{\mathrm{DL}}(a_n)e^{-a_ns}&=\frac{1}{2\pi}\int_0^{2\pi}\frac{1}{1-2\sum_{j=1}^\infty\sum_{m=1}^\infty e^{-sy_m^j\sqrt{p_j}}\cos(k_m^j\theta)}\mathrm{d}\theta\\
&=\frac{1}{2\pi}\int_0^{2\pi}\frac{1}{\left(1-2\sum_{k=1}^\infty e^{-s\sqrt{k(k+2)}}\cos(k\theta)\right)}\mathrm{d}\theta\,,
\end{align*}
which, immediately, gives
\begin{equation}\label{entropy}
  e^{S(a)}=\frac{1}{4\pi^2i}\int_0^{2\pi}\int_{x_0-i\infty}^{x_0+i\infty} \frac{e^{as}}{s\left(1-2\sum_{k=1}^\infty e^{-s\sqrt{k(k+2)}}\cos(k\theta)\right)}\mathrm{d}s\,\,\mathrm{d}\theta\,,
\end{equation}
where the integration contour, formally denoted with the limits $x_0-i\infty$ and $x_0+i\infty$ in the $s$-integration, is the straight line, parallel to the imaginary axis, $\{z:\mathrm{Re}(z)=x_0\}$ with $x_0$ larger than the real part of every singularity in the integrand of \eqref{entropy}.

Several comments are in order now:
\begin{itemize}
  \item The expression \eqref{entropy} is \textit{exact}, but it must be remembered that, to compute $S$ for a value of the area spectrum, it is necessary to take a limit from the right.
  \item It is specially useful to study the asymptotic behavior of $S(a)$ in the large-$a$ regime. To this end, it is possible to employ well-known asymptotic techniques. When these are applied to the present case the Bekenstein--Hawking area law is recovered for a particular choice of the Immirzi parameter $\gamma=0.237\cdots$.
  \item From a practical point of view, the best way to get concrete values of the entropy for not-too-large areas is to use \eqref{GeneratingF4}.
  \item In order to simplify the computations  to understand the behavior of the BH entropy,  it is quite common to sidestep the introduction of the projection constraint and work directly with \eqref{GeneratingF2}. This leads to an approximate value of the entropy given by
      \[
       e^{S^*(a)}=\frac{1}{2\pi i}\int_{x_0-i\infty}^{x_0+i\infty} \frac{e^{as}}{s\left(1-2\sum_{k=1}^\infty e^{-s\sqrt{k(k+2)}}\right)}\mathrm{d}s\,,
      \]
      which is somewhat easier to analyze than \eqref{entropy}. As a matter of fact, this expression leads to the Bekenstein--Hawking law for the same value of $\gamma=0.237\cdots$, but with no logarithmic corrections.
  \item A similar approach can be followed to work with some other prescriptions \cite{Ghosh:2006ph,Engle:2009vc}. The main difference from the ones discussed here, as far as the combinatorial problem  to be solved is concerned, lies in the form of the projection constraint that must be implemented. In any case, the procedure is similar to the one discussed here: introduce an extra variable $z$ in the relevant generating functions. The specific way to do this can be found in \cite{Agullo:2010zz}. In these two instances the value of the Immirzi parameter leading to the Bekestein--Hawking law is $\gamma=0.274\cdots$ which differs from the one given above. The respective log corrections are
      \[
      -\frac{1}{2}\log(a/\ell_P^2)\quad (\mathrm{GM})\,,\quad -\frac{3}{2}\log(a/\ell_P^2)\quad (\mathrm{ENP})\,.
      \]
    As mentioned above,  the logarithmic term for the $U(1)$ case is also
      $
      -\frac{1}{2}\log(a/\ell_P^2).
      $
\end{itemize}

\subsection{Features of the black hole degeneracy spectrum}\label{Features}

Very soon after BHs were considered in the context of LQG,
direct numerical investigations unearthed an unexpected regularity in
the behavior of the entropy as a function of area
\cite{Corichi:2006wn,Corichi:2006bs,Agullo:2008eg}; in fact, the
entropy appeared to be quantized much in the way predicted by
Bekenstein and Mukhanov \cite{Bekenstein:1995ju}. One of the first
applications of the combinatorial and number theoretic methods
described above was to check and confirm the above claims. They were
later used to understand the origin of the observed substructure from
first principles.

The main tool for this purpose is to use the \textit{peak counter}
first proposed in \cite{Agullo:2008eg}. The  idea is to find a way
to partition the space of possible configurations of $k$-labels in
such a way that the peaks observed in the BH degeneracy
distribution (see Figure 3 of \cite{Agullo:2008eg}) are isolated. This
can be achieved by introducing functions in the space of BH
configurations (i.e. the different choices of spin labels for the
edges of the spin network state that pierce the horizon) in such a way
that their level sets select those corresponding to the peaks in the
BH degeneracy spectrum.

Given a configuration $\{(k,n_k)\}$ we define the following functions
\[
N:\mathcal{C}\rightarrow \mathbb{N}: \{(k,n_k)\}\rightarrow \sum_k n_k\,,\, K:\mathcal{C}\rightarrow \mathbb{N}: \{(k,n_k)\}\rightarrow \sum_k kn_k\,,\, P=3K+2N\,.
\]
The first one counts the number of edges of the spin network state
that pierce the horizon, the second adds twice the spin labels of
these edges and the third is a simple combination of both of them. The
level sets $\mathcal{P}_p:=P^{-1}(p)$ of the function $P$ provide a
partition of the space of configurations $\mathcal{C}=\cup_{p\in
\mathbb{N}}\mathcal{P}_p$. An interesting observation
\cite{Agullo:2008eg} is that for each $p\in \mathbb{N}$ the level set
$\mathcal{P}_p$ picks configurations that select \textit{a single
peak} in the BH degeneracy spectrum. This is a non-trivial
fact because other conceivable choices---for instance, the functions
$P_{(\alpha,\beta)}:=\alpha K+\beta N$ with
$\alpha,\beta\in\mathbb{N}$---do not provide such a neat partition of
$\mathcal{C}$ (this is discussed in \cite{FernandoBarbero:2011kb}).

The availability of the peak counter $P$ is very helpful to understand
the staircase structure of the BH entropy for microscopic
BHs and its eventual persistence for large horizon areas. This
is so because the function that gives the entropy as a function of the
area can be built as a sum of the contributions of the individual
steps singled out by $P$. In practice this is done by using the
generating function
\begin{equation}\label{genF_peaks}
\widehat{G}^{\mathrm{DL}}(\nu,s):=[z^0]\left(1-\sum_{k=1}^\infty
\nu^{3k+2}(z^k+z^{-k})e^{-s\sqrt{k(k+2)}}\right)^{-1}\,,
\end{equation}
where a new variable $\nu$ has been introduced in such a way that the
inverse Laplace transform of
$[\nu^\ell]\widehat{G}^{\mathrm{DL}}(\nu,s)$ gives \textit{precisely}
the contribution of the $\ell$-th peak to the BH entropy (see
\cite{FernandoBarbero:2011kb} for details).
After a suitable normalization, the peaks of the BH degeneracy
spectrum can be interpreted as probability densities leading to
probability distributions. A remarkable fact is that the values of the
parameters that describe these distributions (the mean and the
variance, but also higher moments) can be extracted from the
generating function \eqref{genF_peaks} in a straightforward way.
Indeed, the expectation value for the $n$-th power of the area $a^n$
associated with the probability distribution defined by the $p$-th
step in the entropy can be computed as
\[
E(a^n|p)=(-1)^n\frac{[\nu^p]\left(\left.\frac{\partial^n}{\partial
s^n}\right|_{s=0}\widehat{G}^{\mathrm{DL}}(\nu,s)\right)}{[\nu^p]\widehat{G}^{\mathrm{DL}}(\nu,0)}\,.
\]
 From this it is possible to get the mean and the variance of the
distribution and find several useful approximations for
the BH entropy. The crudest one only makes use of the mean and
provides a very good approximation for the position of the observed
steps in the entropy when plotted as a function of the horizon area.
If the mean and the variance are used, it is possible to approximate
the steps as Gaussian distributions. These can be used to understand
the fading of the staircase structure as a consequence of the fact
that their width grows linearly with the BH area.

We end with a word of caution. In the thermodynamical limit, the entropy satisfies some smoothness and concavity conditions. These
are essential in order to use the standard formalism of
thermodynamics in which fundamental quantities, such as the
temperature or the pressure, are defined as derivatives of the entropy.
%(expressed as a function of the energy and other extensive propertiesof the system).
As discussed in \cite{BarberoG:2011gvo}, it is important to understand how
the thermodynamic limit changes the results discussed here, which have
been obtained in the microcanonical ensemble.

\section{Entropy DOF and the Immirzi parameter}\la{sec:DOF}

Now that we have presented the technical details for the derivation of the Bekenstein--Hawking area law from the counting of microscopic states of the horizon  as identified within the LQG framework, let us comment on the nature of these DOF and the related issue of the fixation of the  Immirzi parameter.

We reviewed in Sec. \ref{sec:CC} that the IH phase space---before imposition of spherical symmetry---is characterized by the symmetry group \eqref{GH}. As elucidated in \cite{Freidel:2020xyx, Freidel:2020svx}, different formulations of gravity, related by different choices of boundary Lagrangian, can provide different representations of the corner symmetry group. We saw that in the Palatini formulation, which is the standard starting point of the LQG quantization of IHs, the symmetry group  \eqref{GH} is trivially represented, namely all the IH corner charges vanish. It is only at the quantum level that a finite set of these local SU(2) charges are activated through singular excitations of quantum geometry.
 %by spin network states piercing the horizon.
 In addition, also a subset of tangent diffeomorphisms permuting the punctures  are included in the counting. Therefore, from the standard perspective, all the DOF accounting for the BH entropy have a purely quantum origin, while the classical counting for a spherically symmetric IH would naively lead to a zero entropy.

This perspective, however, may seem quite counterintuitive from a statistical mechanics point of view applied to ordinary systems, like an ideal gas, where any quantum degree of freedom has a classical counterpart. In fact, the situation is usually the opposite than the one described above, with the Gibbs entropy (the classical analog of the von Neumann entropy like \eqref{S}) that is often divergent,  since the properties of classical systems are continuous and the number of classical microstates  uncountably infinite; it is only a {\it coarse graining} of the phase space that renders the classical statistical entropy of the system finite. In the quantum theory, this regularization procedure is implemented through the discreteness of the spectrum of the relevant observable, like the energy, defining the ensemble.

We thus see that the interpretation of the BH entropy DOF can be reconciled with this familiar statistical mechanics point of view if we adopt a gravity formulation, like  the ECH one \eqref{TECH2}, where the IH symmetry group is represented non-trivially in the phase space by having an infinite set of non-vanishing charges at the classical and continuum level. In this case then, the crucial question is wether the quantization of such phase space, or equivalently of the IH corner symmetry algebra, can be achieved and the counting based on the new set of quantum numbers still yields an entropy proportional to the area. While entering this terrain is beyond the scope of this Chapter, let us conclude this section with a few observations and remarks about this alternative description of an IH quantum geometry,   in line with the content of the ``Boundary Degrees of Freedom in Loop Quantum Gravity'' Chapter which can shed light on the role of the Immirzi parameter.

It was shown in \cite{Freidel:2020svx, Freidel:2020ayo} that a   regularization procedure of the infinite-dimensional  corner symmetry algebra,  independent  of a choice of bulk discretization, can be introduced, yielding a
 finite-dimensional   coarse-grained subalgebra associated with \eqref{GH}. This allows one to recover a discrete surface area spectrum already at the continuum and semiclassical level, as well as the LQG flux algebra representing one of the main ingredients of the IH quantization we reviewed. Moreover, applying a similar regularization in the time-gauge context,  a notion of infinitesimal diffeomorphism operators corresponding to spatial translations  on the corner was derived in \cite{Freidel:2019ees}. By an appropriate choice of smearing vector fields on $H$ (reflecting the IH boundary conditions), the latter could be understood as the generators of puncture reordering at the quantum level.
Following this strategy, one could then arrive at a picture where the standard LQG counting presented in Sec. \ref{sec:comb} applies as well to a new construction of the IH Hilbert space based on the quantization of a finite-dimensional subalgebra of the IH corner symmetry algebra. This would put the entropy counting in line with the usual treatment of statistical mechanical systems: The infinite number of horizon classical DOF gets regularized by a
discrete  representation for  the choice of  the area element on the IH cross section, which forms a Casimir of its symmetry algebra and thus acts  diagonally on irreducible representations of $G_H$.

The other advantage of this alternative treatment of the IH phase space is the fact that the role of $\gamma$ becomes apparent already at the semi-classical level. In fact, while leaving the classical bulk dynamics unaffected, only in the presence of a non-vanishing $\gamma$ we have access to a non-trivial representation of the internal Lorentz transformations on the  corner phase space. Moreover, $\gamma$ appears as a proportionality constant between the surface area element and the  SU(2) Casimir already at the semi-classical and continuum levels \cite{Freidel:2020svx, Freidel:2020ayo}; this provides  a  pre-quantization evidence of how its numerical value labels unitarily inequivalent Irreps of the corner symmetry group  defining the non-radiative (kinematical) Hilbert space of the theory. From this perspective then, it is clear how $\gamma$  plays a crucial role already at the semi-classical level and the fact that its numerical value needs to be fixed in order to recover the entropy-area law  becomes a natural feature  of the approach that we have followed.

While these considerations can resolve the possible tension caused by  the numerical fixation of $\gamma$, they surely do not disqualify previous attempts to eliminate it or at least alleviate it. Some of these include:

\begin{itemize}

\item Considering the running from the UV
to the IR  of Newton's constant  and the horizon area within the effective expression for the entropy \cite{Jacobson:2007uj}.

\item Introduction of a new parameterization  of the boundary connection  independent from the Immirzi parameter in the bulk \cite{Perez:2010pq, Engle:2011vf}.

\item  Modification of the first law of BH mechanics through the introduction of a new quantum hair associated with the number of punctures \cite{Ghosh:2011fc}.

\item   Construction of a local quantum Rindler horizon generated by the boost Hamiltonian of Lorentzian Spinfoams \cite{Bianchi:2012ui}.

 \item   Analytic continuation to   $\gamma=i$ in order to restore the full spacetime covariance of the Ashtekar connection and  a proper notion of horizon thermality \cite{Frodden:2012dq, BenAchour:2014erw, Pranzetti:2013lma,Bodendorfer:2013hla,Geiller:2014eza}.

 \item     Introduction of an extra holographic degeneracy factor in the IH partition function  associated with the entanglement  of matter DOF near the horizon \cite{Ghosh:2013iwa}.

  \item    Construction of condensate states in the Group Field Theory  formalism, encoding the continuum spherically symmetric quantum geometry of a horizon \cite{Oriti:2015rwa, Oriti:2018qty}.

 \end{itemize}

 In the end, the correctedness of the value of the Immirzi parameter predicted by the standard LQG BH entropy calculation can be addressed in a conclusive manner only through observational tests sensitive to the area gap; for promising steps in this direction within a cosmological setting see \cite{Ashtekar:2016wpi} and the Chapter  ``Loop Quantum Cosmology: Relation Between Theory and Observations''. Alternatively or (hopefully) in addition to this path, one can hope to have at least another independent theoretical model descending as close as possible from the full LQG framework, where the same numerical value is predicted  by demanding a given outcome or value for an observable of physical relevance.
In this regard, it is intriguing to point out that the value of $\gamma=0.274\cdots$ obtained from the SU(2) counting has been predicted from the study of the effective dynamics describing a Schwarzschild BH interior  as derived from a partial gauge fixing of the full loop quantum gravity Hilbert space; in this model  \cite{Alesci:2020zfi}, the physical relevance of the specific numerical value is related to the behaviour of the post-bounce interior geometry, which approaches an asymptotically de Sitter geometry only for that specific value---see
%e.g. \cite{Easson:2001qf} for the relevance of such scenario  in relation to the classical problems of big bang cosmology and
the Chapter: ``Quantum Geometry and Black Holes'' for other constructions of BH interior effective geometries in LQG.

\providecommand{\href}[2]{#2}\begingroup\raggedright\endgroup

\end{document}